\documentclass{kapproc} 
\def\la{\mathrel{\mathchoice {\vcenter{\offinterlineskip\halign{\hfil
$\displaystyle##$\hfil\cr<\cr\sim\cr}}}
{\vcenter{\offinterlineskip\halign{\hfil$\textstyle##$\hfil\cr<\cr\sim\cr}}}
{\vcenter{\offinterlineskip\halign{\hfil$\scriptstyle##$\hfil\cr<\cr\sim\cr}}}
{\vcenter{\offinterlineskip\halign{\hfil$\scriptscriptstyle##$\hfil\cr<\cr\sim\cr}}}}}

\def\sq{\hbox{\rlap{$\sqcap$}$\sqcup$}}
\def\degr{\hbox{$^\circ$}}

\def\utw{\smash{\rlap{\lower5pt\hbox{$\sim$}}}}
\def\udtw{\smash{\rlap{\lower6pt\hbox{$\approx$}}}}

\def\diameter{{\ifmmode\mathchoice
{\ooalign{\hfil\hbox{$\displaystyle/$}\hfil\crcr
{\hbox{$\displaystyle\mathchar"20D$}}}}
{\ooalign{\hfil\hbox{$\textstyle/$}\hfil\crcr
{\hbox{$\textstyle\mathchar"20D$}}}}
{\ooalign{\hfil\hbox{$\scriptstyle/$}\hfil\crcr
{\hbox{$\scriptstyle\mathchar"20D$}}}}
{\ooalign{\hfil\hbox{$\scriptscriptstyle/$}\hfil\crcr
{\hbox{$\scriptscriptstyle\mathchar"20D$}}}}
\else{\ooalign{\hfil/\hfil\crcr\mathhexbox20D}}%
\fi}}
\def\sqbullet{\vrule height .9ex width .8ex depth -.1ex }  

\def\sqr#1#2{{\vcenter{\vbox{\hrule height.#2pt\hbox{\vrule width.#2pt 
        height#1pt \kern#1pt\vrule width.#2pt}\hrule height.#2pt}}}}

\usepackage{psfig}
\usepackage{rotating}
\usepackage{epsfig}

\newcommand{\apj}{Astrophysical Journal} 
 
\newcommand{\apjl}{Astrophysical Journal Letters} 
 
\newcommand{\mnras}{Monthly Notices of the Royal Astronomical Society} 
\newcommand{\aap}{Astronomy and Astrophysics}

\kluwerbib
\begin{document}
\articletitle[]{Clustering evolution\\
 between $z=1$ and today}
\author{Stefanie Phleps\\
and Klaus Meisenheimer}
\begin{abstract}
We present results of an investigation of clustering evolution
of field galaxies between a redshift of $z\sim1$ and the present
epoch. The current analysis relies on a sample of $\sim 14000$
galaxies in two fields of the COMBO~17 survey. The redshift
distribution extends to $z\sim 1$. The amplitude of the three-dimensional
correlation function can be estimated by means of the projected
correlation function $w(r_p)$. The validity of the deprojection was
tested on the Las Campanas Redshift Survey (LCRS).  In a flat
cosmology with non-zero cosmological constant for bright galaxies ($M_B\leq-18$) 
 the clustering growth is proportional to
$(1+z)^{-2}$. However, the measured clustering evolution  
clearly depends on Hubble type. While locally the clustering strength
of early type galaxies is equal to that of the bright galaxies, at high
redshifts they are  much stronger clustered, and thus the clustering has to evolve
much more slowly.
\end{abstract}

\section{COMBO~17}
The COMBO~17 survey ({\bf C}lassifying {\bf O}bjects by
{\bf M}edium-{\bf B}and {\bf O}bservations in {\bf 17} filters,
see Wolf et al. 2002) has imaged $1\sq\degr$ of sky using the Wide Field
Imager at the MPG/ESO 2.2-m telescope on La Silla. Each of the four
fields has a size of $1/4\sq\degr$. 17 optical filters 
filters have been observed, which facilitates a secure 
multicolor classification and redshift determination 
down to $I=23$.  Each galaxy is assigned a redshift and an $SED$ (in terms of
a number, $0\equiv$ E0, $100\equiv$ extreme starburst). The formal errors in
this process depend on magnitude and type of the object and are of the order of
$\sigma_z=0.02$, and $\sigma_{SED}=2$, respectively. They are getting
larger for galaxies which are undergoing strong starbursts. Therefore
we restricted our analysis to galaxies with reliable redshifts and
take only the more quiescent ones ($SED\leq 80$)  into account.
\section{The projected correlation function}
\begin{figure}[h]
\centerline{\psfig{figure=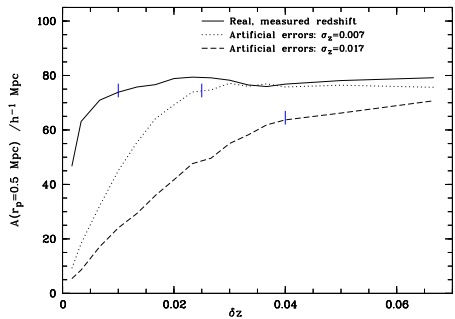,angle=270,clip=t,width=6.5cm}}
\caption[ ]{The influence of redshift measurement errors on the
projected two-point correlation function. $w(r_p)$ for increasing
integration limits is calculated for  $\sigma_z=0.007$ (dotted line), and
for $\sigma_z=0.017$ (dashed line). The errors of the fits are not
plotted here to avoid confusion. The  marks indicate the integration
limits which have to be chosen for the calculation of $w(r_p)$.
\label{wrpmitfehlern}}
\end{figure}

The projected correlation function $w(r_p)$ ($r_p$ is the separation
perpendicular to the line of sight) was first introduced by
Davis and Peebles, 1983, to overcome the influence of non-negligible
peculiar velocities on the three-dimensional correlation
function $\xi(r)$. They show that computing $w(r_p)$ provides a direct measurement of
$\xi(r)$. Redshift inaccuracies, as are unavoidable when using
multicolor redshift estimates, likewise increase the noise and supress
the correlation signal. We calculated $w(r_p)$ for the LCRS
(Shectman et al. 1996), for increasing $\delta z$ (see Figure
\ref{wrpmitfehlern}). The 
amplitude of $w(r_p)$ rises very steeply with 
increasing integration limits, and reaches its maximum at
the point where the integration limits have about the same size as the typical
velocity dispersion in clusters ($\Delta v\approx
2.36\cdot\sigma_v\approx 2500$\,km s$^{-1}$). 
We added artificial errors, to simulate their influence
on $w(r_p)$. As expected, redshift errors shift the point where the plateau is reached
towards larger integration limits. One starts to sample the
correlation signal when the 
integration limits are larger than the {\it Full Width at Half Maximum} of
the redshift error distribution ($\sigma_z=0.007$ corresponds to a
FWHM of $\approx 0.02$, and $\sigma_z=0.017$ corresponds to a FWHM of
$\approx 0.04$, respectively). Note that  the maximum amplitude is also
diminished. For errors of $\sigma_z=0.017$, the maximum amplitude is a factor 1.2
lower than in the case of the unchanged data. Thus we conclude that
the appropriate choice of the integration limits is $\pm\delta z=0.05$. 
\section{The evolution of galaxy clustering}
\begin{figure}[h] \unitlength1cm
\begin{turn}{270}
\begin{picture}(4.5,10)
\put(0,-0.7){\epsfxsize=4.5cm \epsfbox{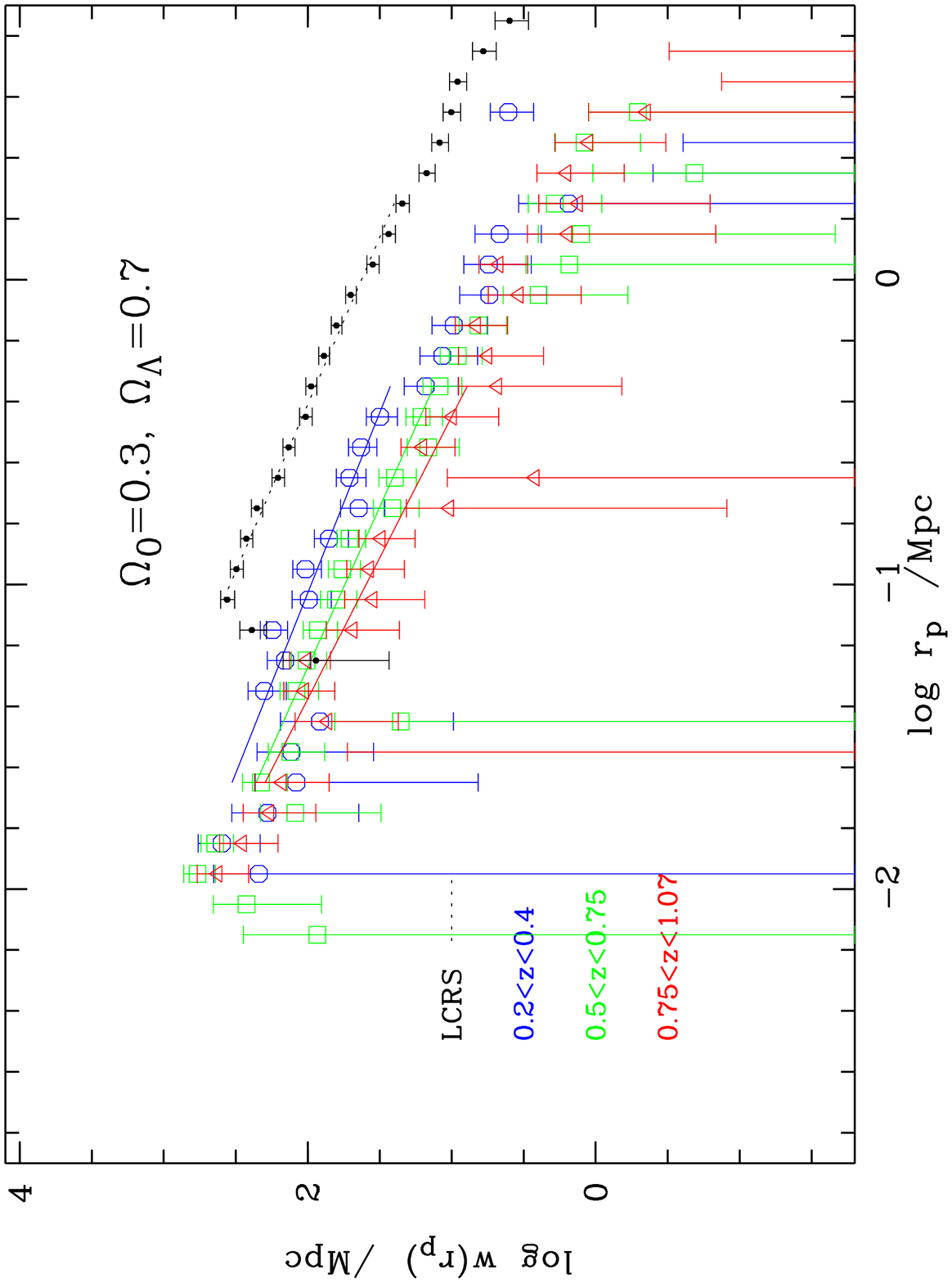}} 
\put(0,6){\epsfxsize=4.5cm \epsfbox{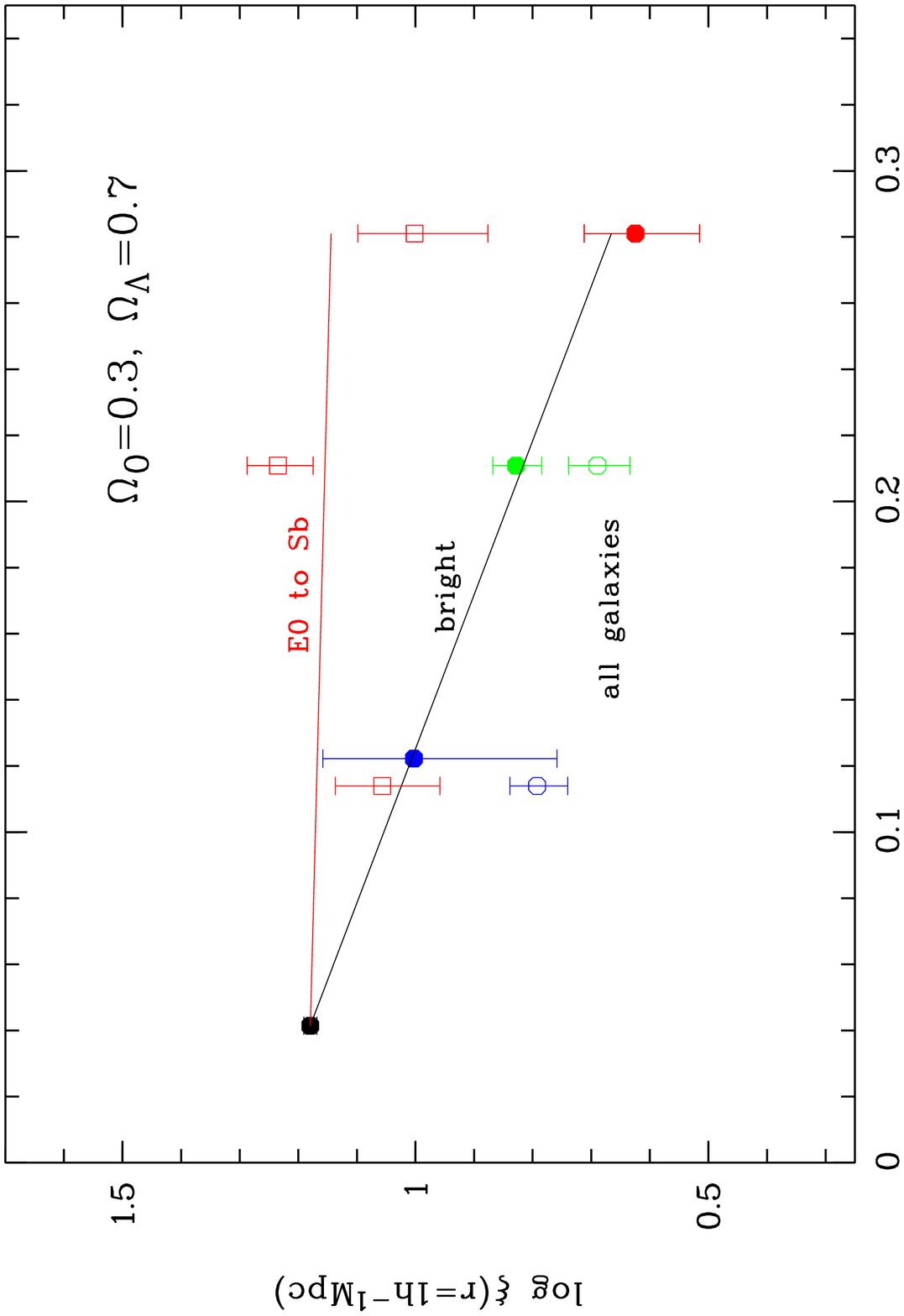}}
\end{picture}
\end{turn}
\caption{Left: Projected correlation function in three redshift bins, for
a flat universe with a non-zero cosmological constant; Right: The
evolution of the clustering strength (at $1 h^{-1}$\,Mpc) with
redshift. The lines are the fits to the data points. The large error
in the bright sample at $z\approx0.3$ is caused by the small
statistics: the calculation for the first redshift bin was started at 
$z=0.25$ instead of $z=0.2$, because there are only very few bright
galaxies at redshifts smaller than 0.3. 
\label{projected}}
\end{figure}

In order to avoid a known galaxy cluster at $z=0.43$ we calculated the
projected correlation function for the COMBO~17 data in 
three different redshift bins ($0.2\leq z<0.4$, $0.5\leq z<0.75$,
$0.75\leq z\leq 1.1$). Although the calculation was carried out for a
closed high-density model ($\Omega_0=1$, $\Omega_\Lambda=0$), a
hyperbolic low-density model ($\Omega_0=0.2$, $\Omega_\Lambda=0$), and
a flat low-density model with non-zero comological constant ($\Omega_0=0.3$,
$\Omega_\Lambda=0.7$), we present here only the latter case (see
Figure \ref{projected}). To facilitate the direct comparison with the 
LCRS data, we used the modified sample (with artificial errors of size
$\sigma_z=0.017$) and the same integration limits.  

Based on the photometry between $350$ and $930$\,nm the absolute
$B$-band luminosities can be estimated without applying any 
$K$-corrections. It is well known that in the local universe bright
galaxies are much stronger clustered than faint ones
(Norberg et al. 2002), as expected if bright (massive) 
galaxies form in the high density peaks of the underlying dark matter
density field (Kaiser 1984). For a reliable
determination of clustering {\it evolution} galaxies with similar absolute
luminosities have to be compared. At redshifts up to $z\la 0.3$ CADIS is
dominated by faint ($M_B <-18$) galaxies, which supress the correlation
signal and thus would distort the measurement of the clustering evolution. A
consistent comparison of galaxy clustering at different redshifts can
only be achieved, if the correlation function in the lowest
two redshift bins is calculated only for galaxies brighter than $M_B=-18$. 

The amplitudes were fitted at $r_p=0.5 h^{-1}\mathrm {Mpc})$ for the
LCRS and for COMBO~17 at $r_p\approx316 h^{-1}$\,kpc (at the mean
redshift of the survey ($\bar{z}$) this corresponds to a comoving separation of
$\approx505h^{-1}$\,kpc, so we compare the same comoving
scales). From $w(r_p)$ and its measured slope we can calculate the
amplitude of $\xi(r)$ at a comoving separation of $r=1
h^{-1}$\,Mpc. Figure \ref{projected} shows the logarithm of the amplitudes of the real
space correlation function at $1 h^{-1}$ \,Mpc, versus $\log
(1+z)$. We parametrise the evolution of the clustering strength with
redshift by a parameter $q$, which gives directly the deviation of the
evolution from the global Hubble flow: $\xi(r_{com}=1{\mathrm
  Mpc})=\xi_0(1+z)^q$, thus the parameter $q$ can be deduced from a 
straight line fit. With a $\Omega_0=0.3$, $\Omega_\Lambda=0.7$
cosmology, we find for the bright sample $q=-2.14\pm 0.20$.

We then repeated the analysis for a subsample of early type galaxies
(E0 to Sb).  The early type galaxies are {\it at higher redshifts}
much higher clustered than the complete sample. At $z=0$ their clustering strengths
are equal, the LCRS point can be used as local measurement for the
early type galaxies as well, and $q=0.15\pm0.24$ 

Field-to-field 
variations do not properly average out, as becomes especially clear in the
early type sample. The full amount of four COMBO~17 is needed to draw
final conclusions concerning the differential 
{\it evolution} of the galaxy clustering. Nevertheless, it is already
obvious that the clustering of the 
older galaxies is significantly different. This is in good agreement
with the concept that biasing should become less and less important
when approaching the present epoch.


\end{document}